\documentstyle[12pt]{article}
\begin{document}
\title{PERTURBATIONS IN BOUNCING COSMOLOGICAL MODELS}
\author{Nelson Pinto Neto\\
CLAFEX/CBPF/CNPq \\ Rua Dr. Xavier Sigaud, 150; 22290-180,
Rio de Janeiro, RJ, Brazil \\ e-mail address: nelsonpn@cbpf.br}
\maketitle

\begin{abstract}
I describe the features and general properties of bouncing
models and the evolution of cosmological perturbations on such
backgrounds. I will outline possible observational
consequences of the existence of a bounce in the primordial
Universe and I will make a comparison of these models with
standard long inflationary scenarios.
\end{abstract}

\section{INTRODUCTION}

The singularity theorems \cite{he} show that, under reasonable physical
assumptions, the Universe has developed an initial singularity,
and will develop future singularities in the form of black holes
and, perhaps, of a big crunch. Until now, singularities are out of the
scope of any physical theory. If we assume that a physical
theory can describe the whole Universe at
every instant, even at its possible moment of creation,
which is the best attitude because it is the only
way to seek  the limits of physical science (in the words of Wheeler
about singularities
\cite{grav}, `physics stops, but physics must go on')
then it is necessary that
the `reasonable physical assumptions'  of the theorems be
not valid under extreme situations of very high energy density and curvature.
We may say that General Relativity (GR), and/or nongravitational field theory,
must be changed under these extreme conditions. In fact, one must view
such theorems as indications of the limits of GR and/or intervention of new
types of matter.

The many possibilities to circumvent the theoerem hypotheses
(quantum effects in geometry and/or matter, non minimal couplings,
curvature squared terms, etc...)  have generated
many cosmological models without singularities which can be classified
in two types:
\vspace{0.3cm}

{\bf 1) Models with a beginning of time.}

As examples belonging to this class, one can cite the ones coming
from euclidean quantum cosmology \cite{euc}, and models tunneling
from nothing \cite{sta}, both in the framework of quantum cosmology.
\vspace{0.3cm}.

{\bf 2) Eternal universes.}

One can divide these models in two subclasses.

{\it i) Always expanding models.}
\vspace{0.1cm}

To avoid the singularity, they must have a phase with $\ddot{a}>0$.
As examples one can cite the pre Big-Bang model \cite{ven} and the
more recent emergent universe \cite{ell}.

{\it ii) Bouncing models.}
\vspace{0.1cm}

In these models, there is a contraction phase preceeding the expansion phase
and a minimal value of the scale factor where $\ddot{a}>0$ {\bf and}
$\dot{a(t_0)}=0$.

Within GR, as we will see in the next section,
realistic bounces without a long inflationary period
after it can occur only for matter contents which violate not
only the SEC but also the Null Energy Condition (NEC,  $\rho + p <0$).
Examples of bouncing models with these properties have as matter
content quantum fields or nonlinear vector fields
described by nonlinear electrodynamics \cite{nov1}
(which can be modelled by two fluids with $p_1=1/3\rho_1$,
$p_2=5/3\rho_2$, with $\rho_2<0$, yielding $a(t)=a_0(1+t^2)^{1/4}$),
radiation plus a negative energy free massless scalar field \cite{bounce}
yielding $a(\eta)=a_0\sqrt{1+\eta ^2}$ ($\eta$ is conformal time).

Ouside GR, bouncing models may be obtained without fluids violating
the NEC condition in models with non minimal couplings \cite{mel},
using Weyl geometries \cite{nov3}, with curvature squared terms
\cite{gur}, and branes within charged ADS black holes \cite{pel}.
Bouncing models can also be obtained in string motivated tree-level actions
\cite{fab} and in quantum cosmology \cite{pin1}.

In this contribution I will describe the features of the bouncing
models and the evolution of cosmological perturbations on them.
In the next section the general properties of bouncing models
will be outlined. In Section III I will describe the evolution
of cosmological perturbations in such backgrounds.
Finally, in the Conclusion, I will outline possible observational
consequences of the existence of a bounce in the primordial
Universe and I will make a comparison of these models with
standard long inflationary scenarios.

\section{PROPERTIES OF BOUNCING MODELS}

Around the bounce, the scale factor behaves generally as

\begin{equation}
\label{sca}
a=a_0 + b\eta ^{2n} + d\eta ^{2n+1} + e\eta ^{2n+2} + ...,
\end{equation}
where $\eta$ is conformal time, $b>0$ and $n\geq 1$.

Defining $\beta\equiv\rho+p$ and $\Upsilon\equiv p'/\rho'$
(which is not necessarily the sound velocity if the fluid violates
the NEC condition), one can write the cosmological equations
for homegeneous and isotropic geometries
in the Einstein frame as

\begin{equation}
\label{f0}
{\cal H}^2 + K  \propto a^2 \rho,
\end{equation}

\begin{equation}
\label{f1}
\beta \propto {\cal H}^2 -{\cal H}'+K,
\end{equation}
where ${\cal H} \equiv a'/a$. Within GR, $\rho$ and $p$ are, respectively,
the energy density and pressure of the matter content of the model.
Outside GR, $\rho$ and $p$ may also depend on gravitational degrees of
freedom.

The characteristics of the fluid in the bounce can be obtained by combining
Eqs. (\ref{sca},\ref{f0},\ref{f1}), yielding (see Ref. \cite{nobounce} for
details)

1) For $n>1$ and $K \neq 0$,
$\Upsilon \propto 1/\eta ^2$ and
$\beta \propto K$

2) For $n>1$ and $K = 0$,
$\Upsilon \propto 1/\eta ^{2n}$ and $\beta <0$

3) For $n=1$, $d\neq 0$ and $\forall K$,
$\Upsilon \propto 1/\eta$ and
$\beta \propto K - 2b/a_0$.

4) For $n=1$, $\forall K$, and $d=0$,
$\Upsilon = {\rm const.}$ and
$\beta \propto K - 2b/a_0$.

For a bouncing model to be realistic, it must be connected to the standard
cosmological model at nucleosynthesis, where $a\propto\eta\propto t^{1/2}$.
In this case, $2b>>a_0$ (see Ref.\cite{nobounce}) and, as one can see from
above, only the fist case with $K=1$ does not violate the NEC condition
in GR.

As in general there is a NEC violation ($\rho+p<0$) and as at nucleosynthesis
the NEC is satisfied ($\rho+p>0$), then there is a conformal time,
$\eta_{\rm NEC}$, called NEC transition time, where $\beta\equiv\rho+p=0$.
At this time one can show that, in all cases,

\begin{equation}
\label{NEC}
\Upsilon\propto 1/(\eta-\eta_{\rm NEC}).
\end{equation}

Hence, in all cases, there is a divergence in $\Upsilon$ either in the
bounce itself or/and in the NEC transition time.

Note that for non interacting fluids in GR where
$p_i = \omega_i \epsilon_i, \qquad \omega_i \nonumber = \hbox{const.}$,
with $\epsilon = \sum_i \epsilon_i$ and $p=\sum_i p_i$ then,
as a consequence of the energy-momentum conservation laws,
$\rho =\rho(a)$ implying that $a(\eta)$ even, with $n=1$ \cite{ent}.
Hence these classes belong to the fourth case above, with
$\Upsilon$ regular at the bounce but fivergent at NEC transition.

\section{THE EVOLUTION OF COSMOLOGICAL PERTURBATIONS ON BOUNCING MODELS.}

The scalar perturbations of homogeneous and isotropic spacetimes
are given by

\begin{eqnarray} \hbox{d}s^2 &=& a^2(\eta) \left\{ (1+2 \phi
)\hbox{d}\eta^2 - 2 B_{;i}\hbox{d}\eta\hbox{d} x^i \right. \nonumber
\\ & & - \left. \left[ (1-2\psi)\gamma_{ij} + 2 E_{;ij}\right]
\hbox{d}x^i \hbox{d}x^j \right\}\; .\label{dg}\end{eqnarray}

Using the gauge-invariant Bardeen potentials are
$\Phi = \phi + [(B-E')a]'/a$ and $\Psi = \psi - a'(B-E')/a$
and the splitting of the gauge invariant pressure perturbation into adiabatic and
entropy components,

\begin{equation}
\delta p = \left({\partial p\over\partial
\epsilon}\right) _S \delta \epsilon+ \left({\partial p\over\partial
S}\right) _\epsilon \delta S = \Upsilon \delta \epsilon + \tau \delta
S,\nonumber
\end{equation}
where $\delta S$ is usually (but not always, specially for
NEC violating fluids) the entropy fluctuation, the perturbed Einstein's
equations yield (we are now restricted to GR)

\begin{eqnarray}
\Phi'' + 3{\cal H} (1+\Upsilon)\Phi' -  \Upsilon \nabla^2 \Phi
+ [ 2{\cal H}'+(1+3\Upsilon)\nonumber \\ \times ({\cal H}^2 -K)]\Phi
= 4\pi Ga^2 \tau \delta S
\label{I}
\end{eqnarray}
\vspace{0.8cm}
Note the presence of $\Upsilon$ in these equations.

For `adiabatic' perturbations, $\delta S = 0$, one obtains

\begin{eqnarray}
\Phi'' + 3{\cal H} (1+\Upsilon)\Phi' -  \Upsilon \nabla^2 \Phi
+ [ 2{\cal H}'+(1+3\Upsilon)\nonumber \\ \times ({\cal H}^2 -K)]\Phi =0.
\nonumber
\end{eqnarray}

As $\Upsilon$ diverges either at the bounce and/or in the NEC
transition, $\Phi$ diverges \cite{nobounce}.
For scale factors behaving as $a=a_0+b\eta^2+e\eta^4...$,
this happens only in the NEC transition.

However as we will now show, this fact does not sign an instability
of these models \cite{ent,nov4}, but simply indicates that entropy
perturbations cannot be neglected at the NEC transition time.

As an explicit example, take two non interacting fluids satisfying
$p_1=\omega_1\rho_1$, $p_2=\omega_2\rho_2$
and define

\begin{equation}
s\propto\left(\frac{\delta \rho_1}{\rho_1(1+\omega_1)} -
\frac{\delta \rho_2}{\rho_2(1+\omega_2)}\right).\nonumber
\end{equation}
One can show, using the perturbed Einstein's equations, that
$s$ and $\Phi$ satisfy

\begin{equation}
\label{eqS27}
s''+{\cal H} (1-3 c_z)s' + k^2 c_z s = \frac{k^2}{\beta} (k^2 - 3K)
\Phi,
\end{equation}

\begin{eqnarray}
\Phi'' + 3{\cal H} (1+\Upsilon)\Phi' -  \Upsilon \nabla^2 \Phi
+ [ 2{\cal H}'+(1+3\Upsilon)\nonumber \\ \times ({\cal H}^2 -K)]\Phi =
\frac{f(a)}{\beta}s,
\end{eqnarray}
where, as before, $\beta = \rho +p$.

Inspection of Eq.~(\ref{eqS27}) shows that entropy
fluctuations cannot be neglected, even for arbitrarily small but
non vanishing values of the wavelength $k$, at the NEC transition
time as long as its source term diverges there. Hence, adiabatic
perturbations cannot be defined at this point and the divergence
detected in Eq. (\ref{I}) for $\delta S = 0$ is not physically meaningful.
In fact, taking the two equations together, one can show that there
are no divergences in the full Bardeen potential \cite{ent}.
Hence, in the NEC transition time, entropy fluctuations are crucial.
In the bounce itself, the same sort of process may happen.

The evolution of cosmological perturbations in bouncing models
has other characteristic features.
In general, bouncing models do not have particle horizons.
Hence, as in inflationary models, one can impose reasonable
physical initial conditions for the perturbations, contrary to the
Standard Cosmological Model without inflation, where initial
conditions are arbitrary due to the lack of physical causal
ineractions when perturbations begin to evolve. In the far past
of general bouncing models the Universe is almost flat. Hence,
one can impose vacuum initial conditions for the perturbations
without any transplanckian
problem based on simple quantum field theory in flat space, yielding
a quantum mechanical origin for them.

In general, after performing suitable changes of variables like
$u=g(a)\Phi$, the perturbation equations can be put in the simple
form $u''+[\Upsilon k^2 - V(a)]u=0$
In power law cosmologies, $V(a)\propto a^2/l_H^2$, where $l_H$ is the
Hubble radius. and hence the transition
from oscillatory regimes ($k^2 > V(a)$) to growing and decaying
regimes ($k^2 < V(a)$) through potential crossing is equivalent to
horizon crossing as long as $k^2 \approx V(a)$ is equivalent to
$\lambda_{\rm phys} \approx l_H$, where the physical wavelength
$\lambda_{\rm phys}$ is given in terms of the comoving wavelength
$\lambda$ through $\lambda_{\rm phys}=a \lambda$ and $k\propto 1/\lambda$.
In bouncing models, it is definitely not true \cite{bounce,jer},
and what is relevant is the potential crossing.

Another important remark is that, defining the transfer matrix as
$\vec{A_+}=${\bf T}$\vec{A_-}$, with $\vec{A}\equiv (A_d(k),A_s(k))$,
$\Phi = A_d(k)f_d(k,\eta)+A_s(k)f_s(k,\eta)$,
where the indices $(+,-)$ refers to (after, before) the bounce and
the indices $(d,s)$ refers to (dominant, subdominant) modes, respectively,
then, contrary to intuition even for very short bounces,
{\bf T} may depend on $k$. A consequence of this is that
matching conditions through a bounce must be treated with care, there are
no general rules, and analysing in detail each particular case is
preferable, withou risks of mistakes.

Bounces can also magnify perturbations which oscillates after it,
and bouncing models can be
constructed in which a scale invariant spectrum for large wavelengths
can be obtained \cite{bounce}.

For gravitational waves, the spectrum is much more complicated then
in inflationary models and highly model dependent, yielding
different possible observational predictions \cite{jer,ekp}.

\section{CONCLUSIONS.}

Bouncing models appear in many instances of physics.
They have no singularities,
no horizon problem, and they may yield a causal explanation and quantum
origin of structures in the Universe,
with possible scale invariant spectrum. If the bounce occurs below
Planck energies, there is no transplanckian problem \cite{tra}.

In realistic models within GR without inflation,
bounces occur only with violation of the NEC.
In the NEC transition time, entropy fluctuations are important and
cannot be neglected ate all.

Many general claims about perturbations on bouncing models,
specially concerning matching conditions
through it, were proven to be erroneous due to counter examples
which have been found \cite{bounce,jer}. Hence, it is safer to
study specific well motivated models.

Some possible observational consequences of a bounce are

a) Oscillations in the primordial spectrum.

b) Different effects in the polarization due to gravitational waves.

These effects taken together may discriminate bounces from
many-fields inflation and transplanckian physics.

Bouncing models may solve issues of inflationary models like
the singularity and transplanckian problem. However, with respect
to initial conditions issues like the flatness problem and
the isotropization problem, bouncing models are silent
\footnote{About the homegeneization problem, both bouncing models
and inflation are silent}. Perhaps
they must be complemented with quantum cosmological ideas, or
be joined with a long inflationary phase after the bounce,
adding the good features of both.
But one must take care: plausible vacuum initial conditions
in the pre bouncing phase may not be transfered
to vacuum initial conditions at the onset of inflation,
a necessary condition to get a Harrison-Zeldovich spectrum.

\section{ACKNOWLEDGEMENTS}

This work was financially supported by CNPq of Brazil.

\end{document}